\documentstyle[epsf,prb,twocolumn,floats,aps]{revtex}
\begin{document}
\twocolumn[\hsize\textwidth\columnwidth\hsize\csname@twocolumnfalse%
\endcsname
\draft
\preprint{IUCM98-015}
\title{Disorder and chain superconductivity in YBa$_2$Cu$_3$O$_{7-\delta}$}
\author{W. A. Atkinson}
\address{Department of Physics, Indiana University, Bloomington IN 47405}
\date{\today}
\maketitle
\begin{abstract}
The effects of chain disorder on superconductivity in
YBa$_2$Cu$_3$O$_{7-\delta}$ are discussed within the context of a
proximity model.  Chain disorder causes both pair-breaking and
localization.  The hybridization of chain and plane wavefunctions
reduces the importance of localization, so that the transport
anisotropy remains large in the presence of a finite fraction $\delta$
of oxygen vacancies.  Penetration depth and specific heat measurements
probe the pair-breaking effects of chain disorder, and are discussed
in detail at the level of the self-consistent T-matrix approximation.
Quantitative agreement with these experiments is found when chain
disorder is present.
\end{abstract}
]

Current understanding of the low energy electronic excitation spectrum
of YBa$_2$Cu$_3$O$_{7-\delta}$ (YBCO) is incomplete.  As with other
high $T_c$ superconductors, YBCO is a layered compound in which the
CuO$_2$ layers are conducting, strongly correlated
quasi-two-dimensional electron gasses.  The CuO$_2$ layers have been
extensively studied\cite{ARPES1} in Bi$_2$Sr$_2$CaCu$_2$O$_8$ using
angle resolved photoemission.  While in-plane conductivity
measurements\cite{Ginsberg} suggest strong similarities between the
CuO$_2$ planes in Bi$_2$Sr$_2$CaCu$_2$O$_8$ and YBCO, a complete
description of YBCO is complicated by the presence of conducting
one-dimensional CuO chain layers, about which relatively little is
known.

Since the CuO chains are unique to the YBaCuO family of
superconductor, one expects to find their signature in a variety of
experiments.  Tunneling experiments,\cite{Dynes} for example, reveal a
finite density of states (DOS) at the Fermi energy in the
superconducting state, in contrast to Bi$_2$Sr$_2$CaCu$_2$O$_8$ where
a clear $d$-wave-like gap is observed.\cite{Renner} Measurements of
the penetration depth $\lambda_c$ along the $c$-axis (perpendicular to
the layers) find a power law dependence on temperature which is
material dependent.\cite{Bonn2,Sridhar,Cooper} More dramatically, a
well developed pseudogap\cite{basov2} is found in the $c$-axis optical
conductivity of YBa$_2$Cu$_3$O$_{7-\delta}$\cite{homes} and
YBa$_2$Cu$_4$O$_8$,\cite{basov2} but not in
La$_{2-x}$Sr$_{x}$CuO$_4$.\cite{basov2,uchida} In some cases, such as
in-plane anisotropic conductivity measurements,\cite{Ginsberg} it is
straightforward to distinguish the contributions of the chains from
those of the planes.  In general, however, the influence of the chains
is not trivial to understand.  Calculations based on multi-band models
suggest, for example, that interband transitions between the plane and
chain bands dominate the $c$-axis conductivity,\cite{Atkinson} and
that the pseudogap seen in optical conductivity experiments reflects a
shift in interband transition energies due to the opening of a
gap\cite{ARPES} in the CuO$_2$-layer Fermi surface.  These predictions
are quite different from those of one-band
models,\cite{Graf,Hirschfeld,Levin} in which disorder and inelastic
scattering are assumed to play a key role in $c$-axis transport.  For
this reason, it is essential to develop a model which correctly
describes the low energy physics of the chain-plane system.

Evidence concerning the chain electronic structure is indirect.  In
YBa$_2$Cu$_3$O$_{7-\delta}$, the chains are not continuous, but are
broken into segments of finite length by the fraction $\delta$ of
vacant chain oxygen sites.  In spite of this, there is a large
anisotropy in the in-plane conductivity,\cite{Ginsberg} indicating
that the chains are metallic.  The absence of localization suggests
that electronic states associated with the chains, while highly
anisotropic, are not one-dimensional.  In the superconducting state,
the in-plane anisotropy\cite{basov} in the penetration depth $\lambda$
is nearly identical in magnitude to the conductivity anisotropy,
suggesting a significant superfluid density on the chain layer for
temperatures $T \ll T_c$.  What is truly remarkable, however, is that
the temperature dependence of the chain superfluid density---as
measured in penetration depth experiments\cite{Hardy}---is almost the
same as that of the planes.  The apparent similarity of the excitation
spectra in the chain and plane layers is surprising given that the
underlying bands have completely different structures.

Several different models have been proposed to describe chain
superconductivity.  The simplest model consistent with $d$-wave chain
superconductivity is the proximity model, in which the pairing
interaction resides in the CuO$_2$ planes and chain superconductivity
occurs through the hybridization of plane and chain wavefunctions.
The failure of this model (discussed below) to describe penetration
depth experiments,\cite{atk1} has led several
authors\cite{wheatley,combescot,Odonovan,Atkinson} to abandon the
premise of a pairing interaction contained exclusively within the
CuO$_2$ planes.

In this work, we show that a small amount of chain disorder is
sufficient to reconcile the proximity model with experiments.  We
model a CuO$_2$-CuO-CuO$_2$ trilayer with a three-band tight-binding
Hamiltonian in which the isolated chain and plane layers have
one and two-dimensional dispersions respectively, 
coupled through single electron hopping.  The resultant bands are
three-dimensional hybrids of chain and plane states.  Because the
hybridization is weak for some values of the in-plane momentum ${\bf
k}$, localization effects are present, but are not sufficient to
eliminate quasiparticle transport in the chains.  We discuss disorder
effects in chain superconductivity in the context of two different
experimental probes of the low energy DOS: specific heat
measurements,\cite{Moler} and penetration depth anisotropy
measurements.\cite{Hardy}

We consider a single trilayer with periodic boundary conditions along
the $c$-axis.  The annihilation operator for an electron in layer $i$
with two-dimensional wavevector ${\bf k}$ and spin $\sigma$ is
$c_{i{\bf k}\sigma}$.  The mean field Hamiltonian for such a model is
(using the Nambu spinor notation)
%\begin{equation} 
${\cal H}
= \sum_{\bf k} C^\dagger_{\bf k} {\bf H}_{\bf k} C_{\bf k} 
$
%\end{equation} 
with
\begin{equation} 
{\bf H}_{\bf k} = \left [
\begin{array}{cccccc} \xi_{1{\bf k}} & t_{\perp 1} & t_{\perp 2} &
\Delta_{\bf k} & 0 & 0 \\ t_{\perp 1} & \xi_{1{\bf k}} & t_{\perp 2} &
0 & \Delta_{\bf k} & 0 \\ t_{\perp 2} & t_{\perp 2} & \xi_{2{\bf k}} &
0 & 0 & 0 \\ \Delta_{\bf k} & 0 & 0 & -\xi_{1-{\bf k}} & -t_{\perp 1}
& -t_{\perp 2} \\ 0 & \Delta_{\bf k} & 0 & -t_{\perp 1} & -\xi_{1-{\bf
k}} & -t_{\perp 2} \\ 0 & 0 & 0 & -t_{\perp 2} & -t_{\perp 2} &
-\xi_{2-{\bf k}}
\end{array} \right ]
\label{1}
\end{equation}
and $C^\dagger_{\bf k} = \left[ c^\dagger_{1{\bf k}\uparrow}
c^\dagger_{2{\bf k}\uparrow} c^\dagger_{3{\bf k}\uparrow} c_{1-{\bf
k}\downarrow} c_{2-{\bf k}\downarrow} c_{3-{\bf k}\downarrow}
\right].$ The hopping amplitude $t_{\perp 1}$ describes hopping
between adjacent CuO$_2$ planes and $t_{\perp 2}$ describes hopping
between plane and chain layers.  The dispersions $\xi_{1{\bf k}}$ and
$\xi_{2{\bf k}}$ describe the isolated plane and chain layers
respectively, while $\Delta_{\bf k}$ is the superconducting order
parameter for the plane layer, which is taken to have $d$-wave
symmetry.  It has been suggested\cite{Andersen} that a suitable tight
binding model for the plane band is
\begin{eqnarray}
\xi_{1{\bf k}} &=& -2t_1[\cos k_xa + \cos k_ya + 2t^\prime\cos
k_xa\cos k_ya \nonumber \\ && + t^{\prime\prime}(\cos 2k_xa + \cos
2k_ya)] - \mu_1
\end{eqnarray}
with $t^\prime = -0.2$ and $t^{\prime\prime} = 0.25$, and $a \approx
3\AA$ the lattice constant.  The chain layer is modeled by
\begin{equation} 
\xi_{2{\bf k}} = -2t_2\cos k_ya - \mu_2.
\end{equation} 
The parameters $t_1$ and $t_2$ can be determined with some certainty
from the magnitudes of the $a$ (in-plane, perpendicular to chains) and
$b$-axis (parallel to chains) penetration depths.\cite{basov} The
other parameters are not as easily determined, but are constrained by
requiring that the Fermi surfaces be consistent with those of
band-structure calculations.\cite{Andersen} Ultimately, our
conclusions are not sensitive to the choice of parameters provided the
above constraints are met.  We take, as plausible:
$\{t_1,t_2,\mu_1,\mu_2, t_{\perp 1},t_{\perp 2}\} =
\{60,400,-20,-600,40,60\} \mbox{ meV}$.  It should be emphasized here
that the chain bandwidth, $4t_2$, determined from $\lambda_b(T=0)$ is
very near to that predicted by band-structure
calculations,\cite{Andersen} consistent with the findings of positron
annihilation studies.\cite{positron}

The order parameter is phenomenological, with
\begin{equation}
\Delta_{\bf k} = \Delta_0(T)[g(k_x) - g(k_y)],
\label{4}
\end{equation}
where the temperature dependence is given by\cite{Rickayzen} $
{\Delta_0(T)}/{\Delta_0(0)} = \mbox{tanh}\left[ {T_c \Delta_0(T)}/{T
\Delta_0(0)} \right ]$.  Based on the location of the van Hove
singularities in the tunneling DOS,\cite{Dynes} we estimate $\Delta_0
= 11 \mbox{ meV}$.  Furthermore, we find that taking $g(k) = \cos(ka)
- 0.3\cos(3ka)$ gives approximately the correct slope\cite{Hardy} for
$\lambda_a(T)$ at low $T$.

\begin{figure}[tb]
\leavevmode
\epsfxsize 0.9\columnwidth
\epsffile{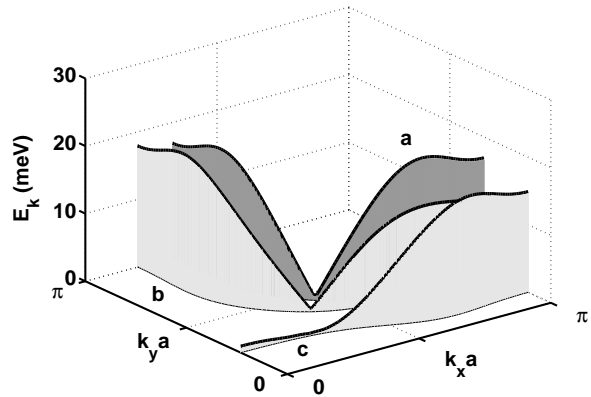}
\caption{Quasiparticle excitation energy $E_{\bf k}$ along the
normal-state Fermi surfaces.  The spectrum reflects both the $d$-wave
structure of the order parameter, and the structure of the underlying
band.}
\label{f1}
\end{figure}

As we shall see, band structure plays a central role in determining
the effects of chain disorder.  The bands (labeled {\bf a}, {\bf b}
and {\bf c}) have dispersions given by the positive eigenvalues of
${\bf H}_{\bf k}$.  In the normal state, $\epsilon^a_{\bf k} =
\xi^-_{\bf k}$, $\epsilon^b_{\bf k} = \epsilon^-_{\bf k}$, and
$\epsilon^c_{\bf k} = \epsilon^+_{\bf k}$ where
\begin{equation} 
\epsilon^{\pm}_{\bf k} = \frac{\xi^+ + \xi_2}{2} \pm \sqrt{\left[
\frac{\xi^+ - \xi_2}{2}\right]^2 + 2t_{\perp2}^2},
\label{7}
\end{equation}
and where $\xi^{\pm}_{\bf k} = \xi_{1{\bf k}} \pm t_{\perp 1}$ are the
energies of bonding and antibonding combinations of the two planes.
From Eq.~(\ref{7}), it is clear that only the antibonding band mixes
with the chain band, and that the bonding band (band {\bf a}) is
completely plane-like.  In Fig.~\ref{f1}, we show the ${\bf
k}$-dependent gap in the superconducting excitation spectrum, plotted
along the Fermi surfaces of the normal state bands.  Band {\bf a} has
the gap structure expected for a $d$-wave superconductor in a
tetragonal system.  Bands {\bf b} and {\bf c} have a more complicated
spectrum, reflecting the structure of the underlying bands.

It is simplest to start with a discussion of impurity effects in the
normal state.  Band {\bf a} is unaffected by chain disorder, and we
focus our attention on bands {\bf b} and {\bf c}.  First, we emphasize
that the behaviour of the two bands, given by Eq.~(\ref{7}), depends
on the degree of chain-plane hybridization, as characterized
by 
\begin{equation} 
\alpha_{\bf k} = \frac{\sqrt{2}t_{\perp2}}{|\xi_{2{\bf k}} -
\xi^+_{\bf k}|}.
\end{equation} 
In the limit $\alpha_{\bf k} \gg 1$, the bands are degenerate relative
to the coupling parameter $t_{\perp2}$ and $\epsilon_{\pm} \approx
(\xi_{2{\bf k}} + \xi^+_{\bf k})/2 \pm \sqrt{2}t_{\perp2}$.  This is
the standard result for the level repulsion of a degenerate two-level
system.  The wavefunctions in this limit are even and odd combinations
of the antibonding and chain wavefunctions, and the electron tunnels
between the bands with a frequency
\begin{equation}
\hbar/\tau_{\bf k} = 2\sqrt{2}t_{\perp2} \quad (\alpha_{\bf k} \gg 1).
\end{equation}  
On time scales longer than $\tau_{\bf k}$, bands {\bf b} and {\bf c}
are three-dimensional.  In the limit $\alpha_{\bf k} \ll 1$, on the
other hand, $\epsilon^b_{\bf k} \approx \min(\xi_{2{\bf k}},\xi^+_{\bf
k}) - \alpha_{\bf k} \sqrt{2}t_{\perp2}$, $\epsilon^{c}_{\bf k}
\approx \max(\xi_{2{\bf k}},\xi^+_{\bf k}) + \alpha_{\bf k}
\sqrt{2}t_{\perp2}$, and the wavefunctions for bands {\bf b} and {\bf
c} are predominantly plane or chain-like.  The tail of band {\bf c}
near $k_x = 0$, for example, is described by this limit.  This weak
mixing of chain and antibonding bands leads to a tunneling rate
\begin{equation}
\hbar/\tau_{\bf k} \approx \sqrt{2}t_{\perp2}\alpha_{\bf k}, \quad
(\alpha_{\bf k} \ll 1)
\label{9a}
\end{equation}
which is much smaller than that of the degenerate case.  In our
trilayer model, $\alpha_{\bf k}$ is strongly ${\bf k}$-dependent due
to the different structures of the (one-dimensional) chain and
(two-dimensional) plane dispersions.  

Because of the large variation in the tunneling rate $\hbar/\tau_{\bf
k}$, the effects of chain disorder are ${\bf k}$-dependent.  If the
average length of unbroken chain is $l$, then we can estimate the
scattering rate to be $\tau_{tr}^{-1} = v_F/l$, where $v_F$ is the
Fermi velocity in the chain layer.  States for which the tunneling
rate $\tau_{\bf k}^{-1}$ between layers is much larger than the
impurity scattering rate $\tau_{tr}^{-1}$ are considered
three-dimensional, and are not susceptible to localization.  On the
other hand, chain-like states for which $\tau_{{\bf k}}^{-1} <
\tau_{tr}^{-1}$ are susceptible to localization.  In our trilayer
model, only a small fraction of chain electrons are localized by chain
disorder.  This observation provides a natural explanation for the
large conductivity anisotropy\cite{Ginsberg} seen in oxygen deficient
YBCO.

\begin{figure}[tb]
\leavevmode
\epsfxsize 0.9\columnwidth
\epsffile{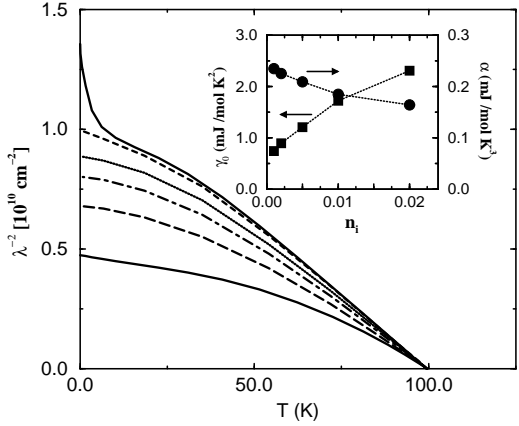}
\caption{Temperature dependence of the penetration depth.  Solid
curves are $\lambda_b^{-2}(T)$ (upper curve) and $\lambda_a^{-2}(T)$
(lower curve) for $n_i = 0$.  Other curves are for $\lambda_b^{-2}(T)$
for impurity concentrations $n_i = 0.001$, $n_i = 0.005$, $n_i =
0.01$, and $n_i = 0.02$ from top to bottom. $\lambda_a^{-2}(T)$ is
almost unchanged by chain disorder.  {\em Inset:} Dependence of the
coefficients of the low temperature specific heat on $n_i$.}
\label{f2}
\end{figure}

We now turn our attention to the superconducting state.  In
Fig.~\ref{f1} we show the ${\bf k}$-dependent superconducting gap
(defined as the energy required to excite a quasiparticle at the Fermi
surface with wavevector ${\bf k}$) produced by the pairing interaction
in the CuO$_2$ plane layers.  The low energy tail in band {\bf c} near
$k_x = 0$ is of particular importance for the current discussion.
States in this part of the band are predominantly chain-like, and the
gap is
\begin{equation} 
E^c_{\bf k} \approx |\Delta_{\bf k}| \alpha_{\bf k}^2 
\quad (\alpha_{\bf k} \ll 1).
\label{10a}
\end{equation}
These low lying states have a significant influence on the low $T$
properties of the system.  In Fig.~\ref{f2} the clean-limit
temperature dependent penetration depth has a pronounced upturn in
$\lambda_b^{-2}(T)$ at low temperatures.  This reflects a sudden
increase in the chain-layer superfluid density as $T$ is lowered
through $E^c_{\bf k}$.

The low temperature upturn in $\lambda_b^{-2}$ is a generic
feature\cite{atk1,wheatley} of proximity models of
YBa$_2$Cu$_3$O$_{7-\delta}$, although the onset temperature for the
upturn depends on the details of the model.  In this work, we show
that small amounts of chain disorder eliminate the upturn.  This is
easily anticipated, since it is apparent that the states which have a
small superconducting gap are chain-like, and are therefore strongly
affected by chain disorder.  The effects of disorder are two-fold:
localization (discussed above) and pair-breaking.  Pair-breaking
occurs for chain-like states whose binding energy $E^c_{\bf k}$ is
less than the scattering rate $\hbar/\tau_{tr}$.  Chain-like
states are characterized by $\alpha_{\bf k} \ll 1$, and in this limit
$E^c_{\bf k}$ is always smaller than the tunneling rate
$\hbar/\tau_{\bf k}$ (assuming that $\Delta_{\bf k} <
\sqrt{2}t_{\perp2}$).  This means that any Cooper pair which satisfies
the criterion $\tau^{-1}_{tr} > \tau_{\bf k}^{-1}$ for localization,
is also broken by disorder.

Localization effects will therefore be important in probes of
quasiparticle transport.  However, since the penetration depth is a
measure of d.c.\ superfluid transport, the important effect of chain
disorder is pair-breaking, which in this work is treated in the
unitary limit of the self-consistent T-matrix approximation.  The
calculation is standard,\cite{Tmatrix} except for the fact that the
impurities reside only in the chain layer.  In the Nambu notation, the
self-energy ${\bf \Sigma}(\omega)$ is a sparse $6\times 6$ matrix with
nonzero elements connecting states in the chain layer only.  
The penetration depth then follows from 
\begin{eqnarray}
\lambda_\mu^{-2}(0) - \lambda_\mu^{-2}(T) &=& -\frac{4\pi e^2}{c^2}
\frac{1}{\beta}\sum_n\frac{1}{\Omega}\sum_{\bf k} \mbox{ Tr}\big[ {\bf
G}({\bf k},i\omega_n) \nonumber \\ && \times {\bf \gamma}_\mu({\bf k})
{\bf G}({\bf k},i\omega_n) {\bf \gamma}_\mu({\bf k}) \big],
\end{eqnarray}
where Tr is the trace over the $6\times 6$ matrix contained in the
square brackets, ${\bf G}({\bf k},i\omega_n) = [i\omega_n - {\bf
H}_{\bf k} - {\bf \Sigma}(i\omega_n)]^{-1}$, and ${\gamma}_\mu({\bf k}) =
\hbar^{-1}\partial {\bf H}_{\bf k} /\partial k_\mu$.  We emphasize that while
this approach is reasonable for making predictions of superfluid and
normal fluid densities, it cannot describe the enhanced backscattering
leading to localization of the normal fluid.\cite{Gogolin}

In Fig.~\ref{f2} we show the effect of disorder on $\lambda_b^{-2}(T)$
for different concentrations $n_i$ of unitary scatterers.  We see that
even small amounts of disorder are sufficient to eliminate the upturn
at low $T$.  Chain disorder, on the other hand, has almost no effect
on $\lambda_a^{-2}(T)$.  In YBa$_2$Cu$_3$O$_{7-\delta}$, the
scattering results from a fraction $\delta$ of chain oxygen sites
which are vacant.  Unfortunately, one cannot simply equate $\delta$
with the fractional impurity density $n_i$ introduced below because of
the tendency of oxygen vacancies to cluster.  Instead, it is more
useful to consider the average length of undamaged chain $l$ (where $l
= a/n_i$) as the fundamental measure of chain disorder.

A useful, and complementary, test of the model comes from examining
the low $T$ specific heat $C_v$, which provides a direct measure of
the quasiparticle density. For $d$-wave superconductors, the
electronic contribution to $C_v$ is well fitted by
\begin{equation}
C_v = \gamma_0 T + \alpha T^2
\end{equation}
where $\gamma_0$ is proportional to the normal fluid density, and
$\alpha$ depends on the ${\bf k}$ dependence of $\Delta_{\bf k}$ near
the gap nodes.  We evaluate $\gamma_0$ and $\alpha$ by fitting the DOS
near the Fermi surface to $N(\omega) = N_0 + N_1 |\omega|$, and then
evaluating $\gamma_0 = N_0 N_a k_B^2 \pi^2/6 \mbox{ mJ/mol/K$^2$}$,
and $\alpha = 9 N_1 N_a k_B^3 \zeta(3) \mbox{ mJ/mol/K$^3$}$, where
$N_a$ is Avogadro's number, $k_B$ is Boltzmann's constant, and
$\zeta(n)$ is the Riemann $\zeta$-function.  In the inset in
Fig.~\ref{f2} we show the effect of chain disorder on both $\gamma_0$
and $\alpha$.  We see that $\gamma_0$ increases rapidly with chain
disorder, as a result of the increasing normal fluid density.  These
results are in excellent quantitative agreement with experiments on
high quality single crystals,\cite{Moler} where $\gamma_0$ was found
to be 1--3 mJ/mol/K$^2$ for $\delta$ between 0.01 and 0.05.

Finally, we comment that we expect impurities to have little effect in
$c$-axis transport, since the states which carry most of the current
along the $c$-axis are those which are most three-dimensional,
and therefore are least affected by chain disorder.  In a series of
conductivity experiments on Zn doped samples, Wang {\em et
al.}\cite{Wang} have found that while the $a$-$b$ anisotropy is
strongly affected by small amounts of Zn, the $c$-axis conductivity is
almost unchanged.

In summary, we have shown that chain superconductivity in
YBa$_2$Cu$_3$O$_{7-\delta}$ is well described by a proximity model, if
one accounts for the presence of disorder in the chain layer.  Within
such a model, hybridization of the chain and plane layers is strongly
dependent on the in-plane momentum ${\bf k}$, so that while most
electronic states are three-dimensional, a fraction of chain-like
states are quasi-one-dimensional and susceptible to localization.  We
calculate the superfluid density in a self-consistent T-matrix
approximation which captures the pair-breaking effects of disorder,
and find good agreement with both penetration depth\cite{Hardy} and
specific heat\cite{Moler} measurements.

We would like to acknowledge J.P.\ Carbotte, W.A.\ Hardy, D.A.\ Bonn,
S.M.\ Girvin and A.H.\ MacDonald for helpful discussions.  This work
was supported by the Midwest Superconductivity Consortium through
D.O.E.\ grant no.\ DE-FG-02-90ER45427 and by the Natural Sciences and
Engineering Research Council of Canada.


\begin{thebibliography}{abcdefghij}

\bibitem{ARPES1} P. Aebi {\em et al}, Phys. Rev. Lett. {\bf 72}, 2757
(1994); D. S. Marshall {\em et al}, Phys. Rev. B, 12 548 (1995);
H. ding {\em et al.}, Phys. Rev. B {\bf 54}, R9678 (1996); Z.-X. Shen
{\em et al.}, Science {\bf 267}, 343 (1995).

\bibitem{Ginsberg} D. B. Tanner and T. Timusk, in {\em Physical
Properties of High Temperature Superconductors III}, edited by
D.M. Ginsberg (World Scientific, Singapore, 1992), P. 363.

\bibitem{Dynes} M. Gurvitch {\em et al.}, Phys. Rev. Lett. {\bf 63},
1008 (1989); J. M. Valles {\em et al.}, Phys. Rev. B {\bf 44}, 11986
(1991).

\bibitem{Renner} Ch. Renner and O. Fischer, Phys. Rev. B {\bf 51},
9208 (1995).

\bibitem{Bonn2} A. Hosseini {\em et al}, Phys. Rev. Lett. {\bf 81},
1298 (1998).

\bibitem{Sridhar} T. Jacobs {\em et al}, Phys. Rev. Lett. {\bf 75},
4516 (1995).

\bibitem{Cooper} C. Panagopoulos {\em et al}, Phys. Rev. Lett. {\bf 79},
2320 (1997).

\bibitem{basov2} D. N. Basov, H. A. Mook, B. Dabrowski, and 
T. Timusk, Phys. Rev. B {\bf 52}, R13 141 (1995).

\bibitem{homes} C. C. Homes, T. Timusk, D. A. Bonn, R. Liang, and
W.N. Hardy, Physica C {\bf 254}, 265 (1995).

\bibitem{uchida}  S. Uchida, K. Tamasaku, and S. Tajima, Phys. Rev. B 
{\bf 53} 14 558 (1996).

\bibitem{Atkinson} {W. A. Atkinson and J. P. Carbotte}, 
Phys.\ Rev.\ B {\bf 55}, 3230 (1997); 12748 (1997); 14592 (1997).

\bibitem{ARPES} D. S. Marshall {\em et al.}, Phys. Rev. Lett. {\bf
25}, 4841 (1996).

\bibitem{Graf} M. J. Graf {\em et al.}, Phys. Rev. B {\bf 47}, 12 089
(1993); 10 588 (1995).

\bibitem{Hirschfeld} P. J. Hirschfeld, S. M. Quinlan, and D. J. Scalapino,
Phys. Rev. B {\bf 55}, 12 742 (1997).

\bibitem{Levin} A. G. Rojo and K. Levin, Phys. Rev. B {\bf 48},
16 861 (1993);  R. J. Radtke, V. N. Kostur, and K. Levin,
Phys. Rev. B {\bf 53} R522 (1996); R. J. Radtke, K. Levin,
Physica C {\bf 250}, 282 (1995). 

\bibitem{basov} D.N. Basov {\em et al.}, Phys. Rev. Lett. {\bf 74},
598 (1995).

\bibitem{Hardy} W. N. Hardy {\em et al.}, Phys. Rev. Lett. {\bf 70},
3999 (1993).

\bibitem{atk1} W. A. Atkinson and J. P. Carbotte, Phys. Rev. B {\bf 52},
10601 (1995).

\bibitem{wheatley} T. Xiang and J. M. Wheatley, Phys. Rev. Lett.
{\bf 76}, 134 (1996).

\bibitem{combescot} R. Combescot, Phys. Rev. B {\bf 57}, 8632 (1998)

\bibitem{Odonovan} C. O'Donovan and J. P. Carbotte, Phys. Rev. B
{\bf  55}, 1200 (1997).

\bibitem{Moler} Kathryn Moler {\em et al.}, Phys. Rev. B {\bf 55},
3954 (1997); Phys. Rev. Lett. {\bf 73}, 2744 (1994).

\bibitem{Andersen} O.K. Andersen, A.I. Liechtenstein, O. Jepsen,
and F. Paulsen, J. Phys. Chem. Solids {\bf 56}, 1573 (1995).

\bibitem{positron} R. Pankaluoto, A. Bansil, L. C. Smedskjaer, and
P. E. Mignarends, Phys. Rev. B {\bf 50}, 6408 (1994); L. C. Smedskjaer,
A. Bansil, U. Welp, Y. Fang, and K. G. Bailey, Physica C {\bf 192},
259 (1992).

\bibitem{Rickayzen} G. Rickayzen, {\em Theory of Superconductivity}
(Interscience, New York, 1965), p. 178.

\bibitem{Tmatrix} G. D. Mahan, {\em Many-Particle Physics} (Plenum,
New York, 1981), pp. 259-265.

\bibitem{Gogolin} A.A. Gogolin, Physics Reports {\bf 86}, 1 (1982).

\bibitem{Wang} N. L. Wang, S. Tajima, A. I. Rykov, and K. Tomimoto,
Phys. Rev. B {\bf 57}, R11081 (1998).





\end{thebibliography}
\end{document}